\documentclass[twocolumn]{revtex4}

\usepackage{graphicx}
\usepackage{dcolumn}
\usepackage{bm}
\usepackage[colorlinks]{hyperref}

\begin{document}


\title{Simple Derivation of the Lifetime and the 
Distribution of Faces\\ for a Binary Subdivision Model
}

\author{Yukio Hayashi}
\email{yhayashi@jaist.ac.jp}
\affiliation{
Japan Advanced Institute of Science and Technology,\\
Ishikawa, 923-1292, Japan}

\date{\today}

\begin{abstract}
The iterative random subdivision of rectangles is used 
as a generation model of networks in physics, 
computer science, and urban planning.
However, these researches were independent.
We consider some relations in them, 
and derive fundamental 
properties for the average lifetime depending on birth-time 
and the balanced distribution of rectangle faces.
\end{abstract}

%


\maketitle


\section{Introduction}
A mathematical model of networks is useful in physics, 
computer science, urban planning, etc. 
There are many methods for constructing spatial 
networks by applying a growing rule or optimization.
One of the attractive methods is based on 
a recursive geometric growing rule 
for the division of a chosen triangle 
\cite{Zhang08,Zhou05,Zhang06,Doye05}
or for the attachment which aims at a chosen edge
\cite{Wang06,Rozenfeld06,Dorogovtsev02} 
in a random or hierarchical selection. 
In particular, 
the fractal-like networks \cite{Hayashi10} generated 
by iterative subdivision of equilateral triangle or square faces 
are more efficient with shorter link lengths and more suitable 
with lower load for avoiding traffic congestion 
than the state-of-the-art complex networks.
These typical complex networks are geometric growing models
\cite{Zhang08,Wang06,Rozenfeld06,Dorogovtsev02,Zhou05,Zhang06,Doye05} 
and the spatially preferential attachment models 
\cite{Brunet02,Manna02,Manna03} 
with various topological structures 
ranging from river to scale-free geographical networks \cite{Nandi07}.
By contrast, 
the advantages of the fractal-like networks are due to the bounded 
path lengths by the $t$-spanner property \cite{Karavelas01}
and the small degrees of nodes without overloaded hubs \cite{Hayashi10}.
The subdivision of squares \cite{Hayashi10} 
is generalized to the subdivision 
of rectangles into four or two smaller 
faces \cite{Hayashi11,Hayashi13}. 
Such a binary subdivision of rectangle faces 
is related not only to 
a self-organization of networks \cite{Hayashi11,Hayashi13} 
in complex network science 
but also to an object generation in computer graphics, 
e.g. the map L-system \cite{Lindenmayer79} 
for road network generation in urban modeling 
\cite{Kato00,Parish01,Frankhauser10}
and the space partitioning for dungeon generation 
in a role playing game (RPG) \cite{Shaker15}.

In addition, 
the hierarchical structure defined by inclusion relations of faces 
is equivalent to a binary tree. 
Binary tree \cite{Mahmoud86} is a well-known date structure 
for sorting numbers in computer science.
It is assumed that 
the input stream of query is a permutation of the integers 
$1, 2, \ldots, N$, whose orderings are at random
in a general problem setting.
For the search task, 
an integer of the input is inserted at a leaf as the terminal 
node that satisfies the ordering condition in any path starting 
at the root.

In spite of the above potential connections, 
these researches were independent. 
Moreover, 
the theoretical analysises for a random binary tree 
\cite{Hattab01,Fekete04}
are a little difficult and probably unknown except in 
a community for mathematicians.

Thus, in this paper, we aim 
\begin{itemize}
 \item to make the derivation of fundamental properties 
   more easily understandable 
 \item to discuss some relations among the findings in the above 
   different research fields
\end{itemize}
for the iteratively random subdivision.

\section{Binary Subdivision Model}
Let us consider the following subdivision model.

\begin{description}
  \item[Step 0:] Set an initial face of rectangle.
  \item[Step 1:] At each discrete time $t = 1,2, 3, \ldots$, 
    chose a rectangle uniformly at random.
  \item[Step 2:] The chosen rectangle face is divided 
    by a line 
    into smaller two ones which is called as twin faces.
  \item[Step 3:] Until the break of a given condition, 
    return to Step 1 at the next time.
\end{description}
In Step 3, for example, we consider a condition: 
the total number of faces is smaller than a given size.

\begin{figure}[htb]
 \centering
 \includegraphics[height=155mm]{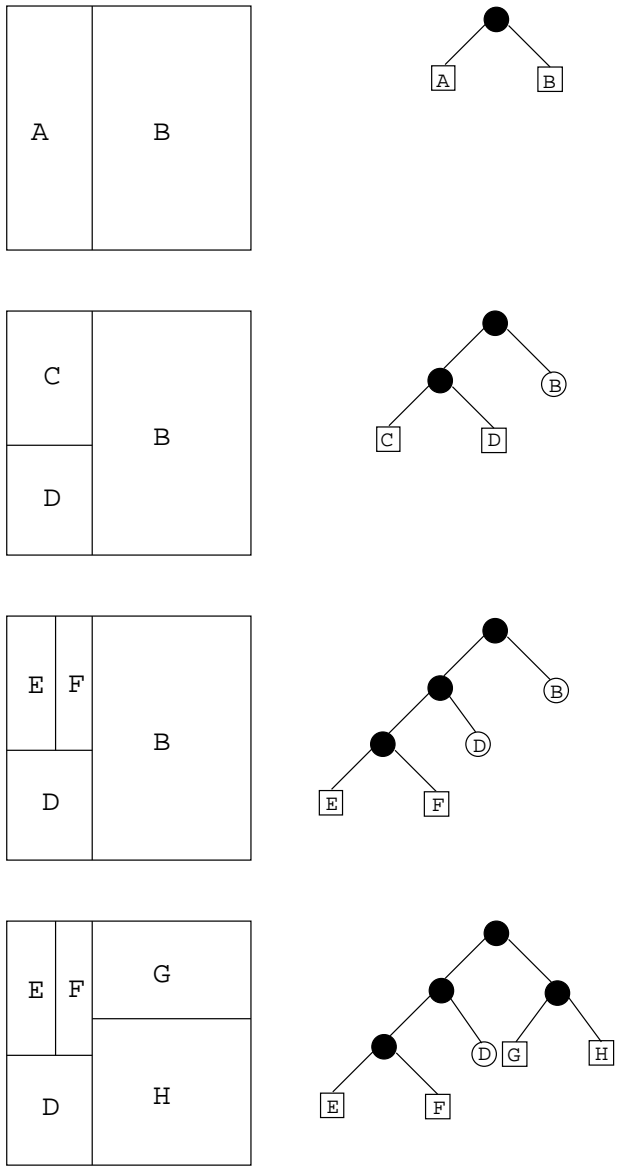}
 \caption{Generation process at $t = 1, 2, 3$ and $4$ 
   from top to bottom. 
   (Left) random subdivision of face. 
   (Right) the corresponding binary tree. 
   The leaf nodes marked by square and open circle represent 
   the twin faces and the other faces, 
   called as feets and arms \cite{Fekete04},
   respectively.
   A, B, $\ldots$, F denotes an identifier of each face.}
\label{fig_division}
\end{figure}

To simplify the discussion without loss of 
the fundamental properties, 
we ignore the area ratio of the divided rectangle 
faces, therefore we do not care the edge lengths of the 
divided rectangle face by 
a bridge line over the chosen face. 
Conceptually, 
the stochastic subdivision of faces is equivalent to 
a random binary tree 
as shown in Fig. \ref{fig_division}, 
although we do not discuss a search problem for random queries.
The leaves in a random binary tree 
represent the rectangle faces, 
which are classified into 
adjacent twin faces generated at a same time and the 
other faces.

\section{Average Lifetime of Faces}
We denote $n_{s}(t)$ as 
the number of rectangle faces counted at time $t$ 
for whose birth-time is $s$. 
In other words, the birth-time is 
the generation time of twin faces by subdivision. 
Although $n_{s}(t)$ is an integer $0, \; 1, \; 2$ in each 
sample of the stochastic process, we consider the average 
behavior over many samples. 
The rate for choosing a face with the birth-time $s$
is proportional to $n_{s}(t)$
because of the uniformly random selection.
Thus, we obtain the expectation 
\[
  n_{s}(t+1) - n_{s}(t) = - \frac{n_{s}(t)}{t+1}, 
\]
and rewrite it to 
\begin{equation}
   n_{s}(t+1) = \left( 1 - \frac{1}{t+1} \right) n_{s}(t), 
   \;\;\; 1 \leq s \leq t,
\label{eq_diff}
\end{equation}
where the total number of faces at time $t$ is 
exactly $t+1$.
The initial configuration is one face, $t=0$.

By recursively applying the difference Eq. (\ref{eq_diff}) with 
the initial condition $n_{s}(s) = 2$, we derive 
\begin{eqnarray}
  n_{s}(t) & = & 
 \Pi_{i=s}^{t-1} \left( 1 - \frac{1}{i+1} \right) \times 2 
\nonumber \\
  & = & \frac{2 s}{s+1} \frac{s+1}{s+2} \ldots \frac{t-2}{t-1} \frac{t-1}{t} 
  = \frac{2 s}{t}.
\label{eq_sol}
\end{eqnarray}
For the average number of faces, 
our new result of Eq. (\ref{eq_sol}) gives 
the time-course of decaying by $1/t$ 
with the dependency on the birth-time $s$.

We consider the expectation time $t'$ 
when the average number of faces becomes one for the 
twin faces with a birth-time $s$. 
Since we obtain $t'=2 s$ 
from $n_{s}(t') = 1$ in Eq. (\ref{eq_sol}), 
the average lifetime of 
more than one face after the births is 
\[
  \Delta t_{s} = t' -s = s.
\]
Thus, younger faces with a larger $s$ 
have a longer lifetime. 
We emphasize that 
this iterative stochastic subdivision is not 
a Poisson process assumed in the analysis for a 
random quadtree model \cite{Eisenstat11}, 
because the selection probability of a face is 
decreasing as time passes 
even with uniform randomness
in increasing the total number of faces. 
Therefore, younger 
faces have less chances for the selection. 
This is not independent and identically distributed.

We consider the case that 
both twin faces generated at time $s$ 
remain at next time $s+1$. 
The probability for the unselection in the total $s+1$ faces 
is 
\begin{equation}
  1 - \frac{2}{s+1} = \frac{s-1}{s+1}.
\label{eq_unselect}
\end{equation}
Until time $t$, 
the remaining probability is given by the product of 
Eq. (\ref{eq_unselect}) 
\begin{eqnarray}
  q_{s}(t) & = & \Pi_{i=s+1}^{t} \frac{i-2}{i} 
\nonumber \\
  & = & \frac{s-1}{s+1} \frac{s}{s+2} 
  \frac{s+1}{s+3} \ldots 
  \frac{t-4}{t-2} \frac{t-3}{t-1} \frac{t-2}{t} 
\nonumber \\
  & = &  \frac{s (s-1)}{t (t-1)}. \nonumber
\end{eqnarray}
By summing $q_{s}(t)$ for all twin faces with birth-times 
$s = 2, 3, \ldots, t-1$, we derive the rate of twin faces 
\begin{equation}
  \sum_{s=2}^{t-1} \frac{s (s-1)}{t (t-1)} 
    = \frac{1}{t (t-1)} \left( \sum_{s=2}^{t-1} s^{2} 
    - \sum_{s=2}^{t-1} s \right) 
    \approx \frac{t}{3}, 
\label{eq_rate_twins}
\end{equation}
where we use the formulas 
$\sum_{s=2}^{t-1} s^{2} = \frac{(t-1) t (2t -1)}{6} -1$ 
and 
$\sum_{s=2}^{t-1} s = \frac{(t-1) t}{2} -1$.
The approximation is valid for a large $t$. 
Since each pair of the twin has two faces, 
the number of faces is $2 t/ 3$ averagely. 
Thus, the rate of other faces is $t / 3$. 
These rates are consistent with the 
asymptotical result \cite{Fekete04} 
derived by a complicated analysis, 
and related to the existing 
rate $n/3$ of nodes with degrees $1$, $2$, and $3$ \cite{Mahmoud86}
in a random binary tree. 
Here, $n = 2t + 1$ denotes the number of nodes 
including leaves with degree $1$, 
non-terminal nodes with degrees $2$ or $3$, 
and a root with degree $2$.
From the rate of these degrees \cite{Mahmoud86}, 
the expected tree has a balanced shape without 
too deep layers by the dominant long chains of node degree $2$.
The balanced tree corresponds to a bell-shape of 
the Poisson distribution as mentioned in the next section.

\section{Distribution of Layered Faces}
\begin{figure}[htb]
 \centering
  \includegraphics[height=40mm]{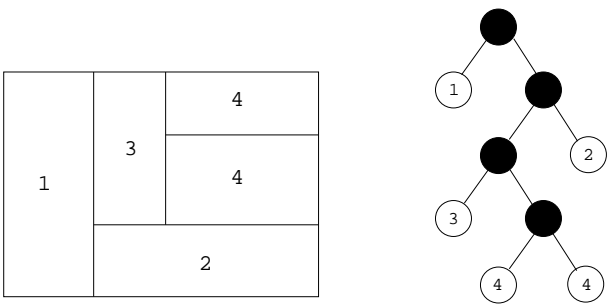}
 \caption{Example of the layer numbered by $l = 1, 2, 3$ and $4$. 
(Left) layered faces.
(Right) the corresponding binary tree.}
\label{fig_layered_faces}
\end{figure}

Next, we consider the distribution of layered faces. 
Faces on the $l$-th layer are one-to-one corresponding to 
the leaves at the depth $l$ in the binary tree. 
The number of faces on the $l$-th layer can increase 
until $2^{l}$. 
Figure \ref{fig_layered_faces} shows an example of 
the layer. 

We denote $N_{l}(\tau)$ 
as the number of faces that belong to the $l$-th layer 
generated after the selections of $l$ times on the 
descendant from the initial face.
As mentioned in the appendix of Ref. \cite{Hayashi11}, 
we consider the random process by subdivision 
in the following continuous-time approximation 
\begin{equation}
  \frac{d N_{l}(\tau)}{d \tau} = 2 \times N_{l-1}(\tau) 
   - N_{l}(\tau), \;\; l \geq 1, \label{eq_IPS1}
\end{equation}
\begin{equation}
  \frac{d N_{0}(\tau)}{d \tau} = - N_{0}(\tau).  \label{eq_IPS2}
\end{equation}
The self-similarity in the iterative subdivision of squares 
\cite{Hayashi11} does not affect 
the analysis of the distribution, because we treat only 
the number of faces on each layer without dependence on the shapes.

The solutions of Eqs. (\ref{eq_IPS1})(\ref{eq_IPS2}) are 
$N_{0} = e^{- \tau}$ and 
\[
   N_{l}(\tau) = 2^{l} \frac{\tau^{l-1}}{(l-1)!} e^{-\tau}, 
   \;\; l \geq 1.
\]
The total number of faces is given by
\[
   {\cal N}(\tau) = \sum_{l} N_{l}(\tau) 
   = 2 e^{\tau},
\]
where we use 
the Taylor series expansion 
$\sum_{l} \frac{(2 \tau)^{l-1}}{(l-1)!} = e^{2 \tau}$. 
Therefore, from $p_{l}(\tau) = N_{l}(\tau)/ {\cal N}(\tau)$, 
we obtain the 
Poisson distribution with a parameter $2 \tau$
\begin{equation}
   p_{l}(\tau) = \frac{(2 \tau)^{l-1}}{(l-1)!} e^{- 2 \tau}, 
   \;\; l \geq 1.
\label{eq_pl}
\end{equation}
Note that 
$p_{l}$ is a function of variable $l$, and $\tau$ is a 
auxiliary variable to take a temporal snapshot. 
The mean and the variance of $p_{l}$ 
follow $2 \tau \propto \log t$ 
by the variable transformation between a linear time variable 
$t$ and a logarithmic time variable $\tau$ 
from the relation for the total number of faces 
$1 + t = 2 e^{\tau}$ \cite{Hayashi11}.
Of course they show the asymptotic behavior for a large $t$. 

Thus, in the bell-shaped Poisson distribution of layers, 
the peak position for the most majority of layers 
shifts to be deeper, 
and the width becomes wider as the divisions are iterated. 
Even if the expanding property can be qualitatively predicted, 
the logarithmic time-course is not trivial.
This simply analyzed property of $\log t$ is also related to 
the deepest level in a random binary tree \cite{Hattab01}.
However, 
the Poisson approximation of distribution was not derived.

\section{Conclusion and Discussion}
By using analytical approaches of difference and differential 
equations, 
we have more easily derived the fundamental properties for 
the average lifetime of faces 
and the distribution of layered faces 
in the iteratively random binary subdivision of rectangles 
which is usually treated as a discrete mathematical problem.
We remark that 
Eqs. (\ref{eq_sol})(\ref{eq_rate_twins})(\ref{eq_pl}) 
hold in more general case 
divided by non-vertical and non-horizontal 
lines with any angles because of no relation to area and shape 
of faces.

Our obtained results will be useful for generating 
road networks in virtual cities \cite{Kato00} 
and dungeons in a RPG \cite{Shaker15}, 
automatically. 
In a dungeon generation, 
the layered faces is applicable to a design of 
corridor placement \cite{Shaker15}. 
In particular, 
rooms assigned to twin faces are connected with a corridor.
Eq. (\ref{eq_rate_twins}) suggests that 
such rooms exist averagely in $2/3$ of the whole rooms 
for the uniformly random divisions, and a game player can directly 
wander back and forth between them. 
When we consider a preferential selection of face 
according to the depth of layer \cite{Eisenstat11} 
instead of the uniformly random selection, 
we can control the rate of twin faces.
The rate becomes larger as a shallow face is chosen for 
the division, then balanced similar depths appear. 
In contrast, 
it becomes smaller as a deeper face is chosen, 
then unbalanced various depths appear. 
So, 
the rate of $2/3$ gives a baseline. 
On the other hand, 
the layered faces represent a historical trace 
of the construction in a road network.
Generally, an area of face is smaller as 
the layer becomes deeper. 
Thus, long-range access roads tend to be constructed at first, 
thereafter 
short-range lanes tend to be added by little and little. 
A bridge lane that produces twin faces by subdivision 
may be related 
to increasing the efficiency of traffic through 
bypaths on the road network.
In the modeling of road networks, we can generate both 
T-shaped and $+$-shaped intersections 
by using a probabilistic selection with a constant mixing rate 
for quartered and binary divisions of faces, 
instead of the uniformly random selection.
Conversely, 
the mixing rate of quartered and binary divisions may be estimated 
from real data of road networks. 
These base line, historical trace, and mixing rate can be also 
discussed in complex network science.

\section*{Acknowledgment}
This research is supported in part by a 
Grant-in-Aid for Scientific Research in Japan, No. 25330100.


\begin{thebibliography}{100}
\bibitem{Zhang08} Z. Zhang, S. Zhou, Z. Su, T. Zou, and J. Guan,
``Random Siepinski network with scale-free small-world and modular
	structure,'' 
{\em Euro. Phys. J. B} Vol.~65, pp.\ 141-147, 2008.

\bibitem{Zhou05}
T. Zhou, G. Yan, and B.-H. Wang,
``Maximal planar networks with large clustering coefficient and
	power-law degree distribution,'' 
{\em Phys. Rev. E} Vol.~71, pp.\ 046141-1-11, 2005.

\bibitem{Zhang06}
Z. Zhang, and L. Rong,
``High dimensional random Apollonian networks,'' 
{\em Physica A} Vol.~364, pp.\ 610-618, 2006.

\bibitem{Doye05}
J.P.K. Doye, and C.P. Massen,
``Self-similar disk packings as model spatial scale-free networks,'' 
{\em Phys. Rev. E}, Vol.~71, pp.\ 016128-1-11, 2005.

\bibitem{Wang06}
L. Wang, F.Du, H.P. Dai, and Y.X. Sun,
``Random pseudofractal scale-free networks with small-world effect,'' 
{\em Eur. Phys. J. B} Vol.~53, pp.\ 361-366, 2006.

\bibitem{Rozenfeld06}
H. D Rozenfeld, S. Havlin, and D. ben-Avraham,
``Fractal and transfractal recursive scale-free nets,'' 
{\em New J. of Phys.} Vol.~6, pp.\ 175-1-15, 2006.

\bibitem{Dorogovtsev02}
S.N. Dorogovtsev, A.V. Goltsev, and J.F.F. Mendes,
``Pseudofractal scale-free web,'' 
{\em Phys. Rev. E} Vol.~65, pp.\ 066122-1-4, 2002.

\bibitem{Hayashi10} Y. Hayashi, and Y. Ono,
``Geographical networks stochastically constructed by a self-similar
	tiling according to population,''
{\em Phys. Rev. E} Vol.~82, pp.\ 016108-1-9, 2010.

\bibitem{Brunet02} R. Xulvi-Brunet, and I.M. Sokolov, 
``Evolving networks with disadvantaged log-range connections,''
{\em Phys. Rev. E}, Vol.~66, pp.\ 026118, 2002.

\bibitem{Manna02} S.S. Manna, P. and Sen, 
``Modulated scale-free network in Euclidean space,''
{\em Phys. Rev. E}, Vol.~66, pp.\ 066114, 2002.

\bibitem{Manna03} P. Sen, and S.S. Manna, 
``Clustering properties of a generalized critical Euclidean network,''  
{\em Phys. Rev. E}, Vol.~68, pp.\ 026104, 2003.

\bibitem{Nandi07} A.K. Nandi, and Manna, 
``A transition from river networks to scale-free networks,'' 
{\em New Journal of Physics}, Vol.|9, pp.\ 30, 2007.

\bibitem{Karavelas01}
M. I. Karavelas, and L. J. Guibas,
``Static and kinetic geometric spanners with applications,''
{\em Proc. of the 12th ACM-SIAM Symposium on Discrete Algorithms},
pp.\ 168--176, 2001.

\bibitem{Hayashi11} Y. Hayashi, 
``An Approximative Calculation of the Fractal Structure in 
Self-Similar Tilings,'' 
{\em IEICE Trans. on Fundamentals}, Vol.~E94-A, No.~2, 
pp.\ 846-849, 2011.

\bibitem{Hayashi13} Y. Hayashi, T. Komaki, Y. Ide, T. Machida, 
and N. Konno, 
``Combinatorial and approximative analyses in a spatially 
random division process,'' 
{\em Physica A}, Vol.~392, pp.\ 2212-2225, 2013.


\bibitem{Fekete04} E. Fekete, 
``Arms and Feet Nodes Level Polynomial in Binary Search Trees,'' 
In M. Drmota, P. Flajolet, D. Gardy, and B. Gittenberger (Eds.), 
Mathematics and Computer Science III
Trends in Mathematics, pp.\ 229-240, 
Birkh\"{a}user, 2004.

\bibitem{Hattab01} J. Jabbour-Hattab, 
``Martingales and Large Deviations for Binary Search Trees,'' 
{\em Random Structure and Algorithm}, Vol.~19, pp.\ 112-127, 2001.

\bibitem{Mahmoud86} H.M. Mahmoud, 
``The Expected Distribution of Degrees in Random Binary Search Trees,'' 
{\em The Computer Journal}, Vol.~29, No.~1, pp.\ 36-37, 1986. 

\bibitem{Eisenstat11} D. Eisenstat, 
``Random road networks: the quadtree model,'' 
{\em Proceeding of SIAM the 8th Workshop on 
Analytic Algorithms and Combinatorics} (ANALCO11), Jan. 22, 2011. 
http://arxiv.org/abs/1008.4916

\bibitem{Lindenmayer79} A. Lindenmayer, and G. Rozenberg, 
``Parallel Generation of Maps: Developmental Systems for Cell Layers,'' 
{\em Lecture Notes in Computer Science}, Vol.~73, pp.\ 301-316, 1979.

\bibitem{Kato00} N. Kato, T. Okubo, H. Kanoh, and S. Nishihara, 
``L-system Approach to Generating Road Networks for Virtual Cities,'' 
{\em IPSJ}, Vol.~41, No.~4, pp.\ 1104-1112, 2000 (in Japanese).

\bibitem{Parish01} Y.I.H. Parish, and P. M\"{u}ller, 
``Procedual Modeling of Cities,'' 
{\em Proceedings of the 28th annual conference 
on Computer graphics and interactive techniques}, 
SIGGRAPH 2001, pp.\ 301-308, 2001.

\bibitem{Frankhauser10} P. Franlhauser, 
``Fractal Geometry for Measuring and Modeling Urban Patterns,'' 
In S. Albeverio, D. Andrey, P. Giordano, and A. Vancheri (Eds.), 
The Dynamics of Complex Urban Systems 
-An Interdisciplinary Approach-, 
pp.\ 213-43, Physica-Verlag Springer, 2010.

\bibitem{Shaker15} N. Shaker, A. Liapis, J. Togelius, R. Lopes, 
and R. Bidara, 
``Constructive generation methods for dungeons and levels (DRAFT),'' 
In N. Shaker, J. Togelius, and M.J. Nelson (Eds.), 
Procedural Content Generation in Games, 
Chapter 3, pp.\ 31-55, 2015. 
http://pcgbook.com/

\end{thebibliography}
\end{document}